\documentclass{article}

\usepackage{arxiv}

\usepackage[utf8]{inputenc} % allow utf-8 input
\usepackage[T1]{fontenc}    % use 8-bit T1 fonts
\usepackage{hyperref}       % hyperlinks
\usepackage{url}            % simple URL typesetting
\usepackage{booktabs}       % professional-quality tables
\usepackage{amsfonts}       % blackboard math symbols
\usepackage{nicefrac}       % compact symbols for 1/2, etc.
\usepackage{microtype}      % microtypography
\usepackage{graphicx}
\usepackage{doi}
\usepackage{amsmath}

\title{Modeling restricted enrollment and optimal cost-efficient design in multicenter clinical trials }

\date{}

\def\Gam{{\rm Ga}}

\def\NB{{\rm NB}}

\def\PG{{\rm PG}}

\def\a{\alpha}
\def\be{\beta}

\def\Ga{\Gamma}

\def\la{\lambda}
\def\La{\Lambda}
\def\si{\sigma}

\def\pto{\stackrel{{\rm P}}{\longrightarrow}}

\def\E{{\mathbf E}}
\def\Pr{{\mathbf P}}
\def\Va{{\mathbf {Var}}}

\def\cN{{\mathcal N}}

\def\wti{\widetilde}

\def\nn{\nonumber}

\newtheorem{lemma}{Lemma}[section]

\def\bee{\begin{equation}\label}
\def\ene{\end{equation}}
\def\beeq{\begin{eqnarray}\label}
\def\eneq{\end{eqnarray}}
\def\beeqn{\begin{eqnarray*}}
\def\eneqn{\end{eqnarray*}}
\def\beth{\begin{theorem}\label}
\def\enth{\end{theorem}}

\def\belem{\begin{lemma}\label}
\def\enlem{\end{lemma}}

\author{
\\
{\large Vladimir Anisimov\thanks{E-mail: \texttt{Vladimir.Anisimov@amgen.com}}}
\\
	Data Science\\
	Center for Design \& Analysis\\
	Amgen, London, UK \\
	\And
	\\ {\large Matthew Austin} \\
	Data Science\\
	Center for Design \& Analysis\\
	Amgen, Thousand Oaks, CA, US \\
}

\renewcommand{\headeright}{}
\renewcommand{\undertitle}{}
\renewcommand{\shorttitle}{Modeling restricted enrollment and optimal cost-efficient design}

\hypersetup{
pdftitle={Modeling restricted enrollment and optimal cost-efficient design in multicenter clinical trials},
pdfauthor={Vladimir Anisimov, Matthew Austin},
pdfkeywords={Patient enrollment, Poisson-gamma model, Forecasting enrollment,
Optimal enrollment design},
}

\begin{document}

\maketitle

\begin{abstract}
Design and forecasting of patient enrollment is among the greatest challenges
that the clinical research enterprize faces today, as inefficient enrollment
can be a major cause of drug development delays.
Therefore, the development of the innovative statistical and artificial intelligence
technologies for improving the efficiency of
clinical trials operation are of the imperative need.
This paper is describing further developments in the
innovative statistical methodology
for modeling and forecasting patient enrollment.
The underlying technique uses a Poisson-gamma enrollment model
developed by Anisimov \& Fedorov in the previous publications
and is extended here to analytic modeling of the enrollment on country/region
level.
A new analytic technique based on the
approximation of the enrollment process in country/region by a
Poisson-gamma process with aggregated parameters is developed.
Another innovative direction is the development of the analytic technique for
modeling the enrollment under some restrictions
(enrollment caps in countries).
Some discussion on using historic trials for better prediction of the enrollment
in the new trials is provided.
These results are used for solving the problem of optimal trial cost-efficient enrollment design:
find an optimal allocation of sites/countries that minimizes the global trial cost
given that the probability to reach an enrollment target in time
is no less than some prescribed probability.
Different techniques to find an optimal solution for high dimensional
optimization problem for the cases of unrestricted and restricted enrollment and
for a small and a large number of countries are discussed.
\end{abstract}

% keywords can be removed
\keywords{Patient enrollment, Poisson-gamma model,
Forecasting enrollment, Restricted enrollment, Optimal enrollment design}

\section{Introduction}\label{sec:1}
The multibillion clinical trials market is in an outstanding need of
using innovative statistical and artificial intelligence
technologies for improving the efficiency of
clinical trials operation as 80\% of clinical trials fail to meet
enrollment timelines.

Statistical design and trial operation are affected by stochasticity
in patient enrollment and various event appearance. The complexity
of clinical trials and multi-state hierarchic structure of different
operational processes require developing new predictive analytic
techniques
for efficient data analysis,
forecasting/monitoring \& optimal decision making.
%for efficient modeling and forecasting trial operation.

There are many challenging problems in trial design.
According to a research from the Tufts Center for the Study of Drug
Development \cite{tufts13}, while 9 out of 10 clinical trials
worldwide meet their patient enrollment goals, reaching those targets
typically means that drug developers need to nearly double their
original timelines.
Citing Ken Getz, director of
sponsored research at Tufts Center for the Study of Drug
Development,
"Patient recruitment and retention are among the greatest challenges
that the clinical research enterprise faces today, and they are a
major cause of drug development delays".

Patient enrollment is one of the main engines driving operation of
contemporary late stage trials.
There are many uncertainties in input data and
randomness in enrollment over time.
Enrollment stage is very costly, it also affects
many other operational characteristics: follow-up stage, supply
chain and time to deliver drug on market.
Many companies still use ad-hoc simplified or deterministic models.
This may lead to inefficient design, underpowered and delayed
trials,  extra costs and drug waste.

The key questions for all pharmaceutical companies and CRO's: How do we improve
predictability of patient enrollment with the goal to improve the
efficiency and quality of clinical trial operation? Which countries
and how many sites should we select for study that: enrol the
fastest with minimal cost to get a desired Probability of Success?

Historically, the main attention of statisticians working in clinical research
is paid to the statistical trial design, sample size analysis, without
giving much consideration to the investigation of the impact of a patient enrollment process
on the whole study operational design.

However, as at any future time point the number of patients at different levels and in
different cohorts are uncertain, to use the proper stochastic models to account
for these uncertainties is a key as this will allow one to predict the times of interim
and final analysis, and evaluate the resources required to reach the trial goals in time.

Nowadays, the late phase clinical trials typically involve hundreds or even thousands of patients
recruited by many clinical sites among different countries.
Some controversies in the analysis of multicenter clinical
trials are considered in \cite{senn97,senn98}.
%by Senn \cite{senn97,senn98}. %(1997,1998).

Therefore, we investigate clinical trials where the patients are
recruited by multiple clinical sites.
At the initial stage of trial design and at the interim stage
the imperative tasks are predicting the number of patients to be recruited
in different countries/regions as this impacts the whole trial operational design.

There is quite extensive literature on using different approaches
for  enrollment modeling.
Quite a large number of papers are devoted to using mixed Poisson models.
In \cite{williford87}
%Williford et al. \cite{williford87} %(1987)
the authors use a Poisson process with gamma distributed rate
to model the global enrollment process.
Several authors
(\cite{senn97,senn98,carter05}) %, 1997, 1998).
%(Carter et al., \cite{carter05}; %2005;
%Senn \cite{senn97,senn98}) %, 1997, 1998).
use the Poisson processes with fixed
recruitment rates to describe the enrollment process in different clinical sites.

However, in real trials different sites typically may have different capacity and productivity,
thus, the enrollment rates in different sites vary. %Therefore,
To reflect this variation, Anisimov and Fedorov
\cite{anfed05,anfed07a,anfed07}
%Anisimov and Fedorov
%(2006, 2007)
introduced a so-called Poisson-gamma model, where the variation in rates
is described using a gamma distribution.
First, this model was  introduced in \cite{anfed05}
where various characteristics of the number of patients in different sites
at the end of the enrollment were investigated.
Then in \cite{anfed07a,anfed07}) this model was used for modeling and predicting
the enrollment processes over time accounting also for different times of sites initiation.
% and for interim re-projection of the enrollment.
This model can be also seen in the framework of the empirical Bayesian approach
where the prior distribution of the rates is a gamma
distribution with the parameters that at the initial stage can be evaluated either using historic data
or expert estimates of study managers.
% (Anisimov \cite{an11a}).
% 2011b).
%Anisimov and Fedorov \cite{anfed07}
In  \cite{anfed07a,anfed07} it was also proposed
a maximum likelihood technique for estimating the parameters of the rates %at any interim time
and the Bayesian technique for adjusting the posterior
distribution of the rates at any interim time using enrollment data in the individual sites.

Later on in \cite{gajew-sim08},
%Gajewski et al. \cite{gajew-sim08}
%(2008, 2012)
it was independently considered a similar model for modeling enrollment
using a Poisson process with gamma distributed rate but assuming that there is only one clinical site.

To capture wider realistic scenarios, the technique based on using
a Poisson-gamma model was developed further to
account for random delays and closure of clinical sites and analysis of some performance
measures \cite{an-dow-fed07,an08,an11a,an20}.
%(Anisimov et al. \cite{an-dow-fed07}, Anisimov \cite{an11a}).
The Poisson-gamma model was  used as a baseline methodology in \cite{an11b,an-ArXiv21}
for modeling event counts in event-driven trials,
%(Anisimov \cite{an11b}),
in \cite{an16a}
for forecasting various trial operational characteristics associated with enrollment,
%(Anisimov \cite{an16a}), and
and in \cite{an-aus20} for centralised statistical monitoring of
clinical trial enrollment performance.
%(Anisimov \& Austin \cite{an-aus20}).

The Poisson-gamma model was also used in \cite{baksenn13}
%by Bakhshi et al. \cite{baksenn13} % (2013)
for evaluating the parameters of the model using meta-analytic techniques of historic trials,
and in \cite{savy12,savy17}
%by Mijoule et al. \cite{savy12,savy17} %(2012)
to investigate the opportunity of using Pareto distribution for the enrollment rates
%and in Minois et al. \cite{savy17}
and for evaluating the duration of recruitment when historic data are available.
A survey on using mixed Poisson models is provided
in a discussion paper \cite{an16b}.
%(Anisimov \cite{an16b}).

Note that a mixed Poisson-gamma distribution and the associated negative binomial
distribution were also used in other applications, e.g. in \cite{bates52}
%by Bates and Neyman \cite{bates52},
%(1952),
for describing the
variation of positive variables in modeling flows of various events.

There are also other approaches to enrollment modeling described in the literature, however,
they are dealing mainly with the analysis of global enrollment
and therefore have some limitations. Specifically, these approaches typically require
rather large number of sites and patients (to use some approximations) and cannot
be applied on the level of site/country for evaluating the enrollment performance and
forecasting.
There are different techniques used and the readers can look at survey papers
\cite{barnard10,gkioni19,heitjan15}
%by Barnard et al. \cite{barnard10}; Heitjan et al. \cite{heitjan15}),
%Gkioni et al. \cite{gkioni19}
and also a discussion paper \cite{an12}.
%(Anisimov \cite{an12}).

The purpose of this paper is to develop further the basic methodology
for analytic modeling of enrollment on different levels,
consider practical cases of upper restrictions on the enrollment on country level
and also propose the techniques for solving the problem
of optimal cost-efficient enrollment design given some cost/timelines constraints
for unrestricted and restricted enrollment, both.

As we need to model the enrollment  on different
levels, the approaches oriented to modeling global enrollment
are not suitable here.
As the baseline model we use a Poisson-gamma enrollment
model
for modeling enrollment on site level.
The enrollment
processes on country/region levels are described by mixed
Poisson processes with some aggregated characteristics
%(Cox processes)
%(Cox and Isham \cite{cox80}),
and depend on the site's initiation and closure.

The paper is organized as follows. Section 1 is devoted to some background
and literature survey. In Section 2 a Poisson-gamma
enrollment model for unrestricted (competitive) enrollment is introduced as these results
are used in the further presentation.
Section 3 is devoted to modeling/predicting
enrollment under upper restrictions on country level
and to the investigation of the impact of enrollment caps.
A brief discussion on using historic data for better predicting enrollment
rates for the new trials is also provided.
Section 4 is devoted to the discussion of different approaches/techniques on how to create
an optimal cost-efficient enrollment design: find an optimal sites/countries allocation that minimizes
the total trial cost given that the probability to complete enrollment in time is no less than some prescribed probability
and there are certain restrictions on the number of sites in countries.
Some results on the approximation of the convolution of Poisson-gamma variables
and on the calculation of the mean and the variance of the restricted enrollment process 
 are given in Appendix.

\section{Enrollment modeling}\label{sec-2}

In this section, in
subsection \ref{sec2-1} we review some basic notation and properties of a
Poisson-gamma enrollment model (referred to as a PG model) that will be used
throughout the paper.
 The presentation here  mainly follows \cite{an11a}.
% (Anisimov \cite{an11a}).
Subsection \ref{sec2-2} presents a novel analytic technique for modeling
enrollment process on country level using the approximation by a Poisson-gamma process.
These results are essential for developing the analytic technique for modeling
enrollment under upper restrictions on country level
which is investigated in the next Section \ref{sec3}.

\subsection{Modeling unrestricted enrollment}\label{sec2-1}

Consider a clinical trial where the patients are recruited by
different clinical sites and, after a screening period, they are
randomized to different treatments.
Most of clinical trials use so-called competitive enrollment (no
restrictions on the number of patients to be recruited in particular
sites/regions). Nevertheless, sometimes due to some geographical
or population reasons, clinical teams may use restricted enrollment,
e.g. in some countries/regions there might be an upper (or lower) threshold
(say, to enrol no more (or no less) than a given number of patients).

In this subsection we consider first a competitive enrollment (no
restrictions).

Assume that the patients arrive to each clinical site one at a
time and independently of each other. Then the natural model to describe
the arrival flow in site $i$ is a Poisson process with some rate $\la_i$.
As the value of the rate may not be certain and can be evaluated
only up to some uncertainties, it is natural to model a variation in the rate
using a gamma distribution.
Moreover, as patients arrive at different sites independently,  we
assume that rates $\la_i$ are jointly independent random variables.

This enrollment model is developed by Anisimov \& Fedorov
and is called a Poisson-gamma (PG) model \cite{anfed07a,anfed07}).
It is also extended further in some directions in \cite{an09a,an11a,an20}.
% in (Anisimov \& Fedorov \cite{anfed07a,anfed07}) and further extended in
%(Anisimov \cite{an09a,an11a,an20}).
%(Anisimov et al. \cite{an-dow-fed07};

Let us introduce some basic notation that will be used throughout
the paper.

Denote by $\Pi_a(t)$ an ordinary homogeneous Poisson process with
rate $a$, so, for any $t>0$,
$$
\Pr(\Pi_a(t) = k) = e^{-at}\frac{(at)^k}{k!}, \ k=0,1,...
$$
where we set $0! = 1$ and $0^0 = 1$.
Denote also by $\Pi(a)$ a Poisson random variable with parameter
$a$.
Let $\Gam(\alpha,\beta)$ be a gamma
distributed random variable with parameters $(\alpha,\beta)$ (shape and rate) and
probability density function
\begin{equation}\label{e00}
f(x,\alpha,\beta) =
\frac {e^{- \beta x} \beta^{\alpha} x ^{\alpha-1} }{ \Gamma(\alpha)},\
x >0,
\end{equation}
where \,
$ \Gamma(\alpha) = \int_0^\infty e^{- x}  x ^{\alpha-1} {\rm d}x$ \, is a gamma function.

Assume now that the rate $\la$ has a gamma distribution
with parameters $(\alpha,\beta)$ and introduce a mixed (doubly stochastic)
Poisson process $\Pi_\la(t)$.
According to
\cite{bernardo04},
$\Pi_\la(t)$ is a Poisson-gamma (PG) process with parameters $(t,\alpha,\beta)$ and
\bee{PG}
\Pr(\Pi_\la(t) =  k)
= \frac{\Gamma(\alpha + k)}{k!\ \Gamma(\alpha)}\ \frac{t^{k}\beta^{\alpha}}
{(\beta + t)^{\alpha + k}}, \ k = 0,1,..
\ene
For convenience denote also by $ \PG(t,\alpha,\beta)$ a PG random variable
 that has the same distribution
as $\Pi_\la(t)$.

For $t = 1$, $\Pi_{\lambda}(1)$ has the same distribution as $\Pi(\la)$ (mixed Poisson variable).
In this case for simplicity we use notation $\PG(\alpha,\beta)$ instead of $ \PG(1,\alpha,\beta)$.

Note that according to (\cite{john-kotz93}, p. 199),
%Johnson et al. (1993, p. 199)
the distribution of $\Pi_{\lambda}(t)$ in (\ref{PG}) can be also described as
a negative binomial distribution, and for any $t >0$,
\bee{NB1}
\Pr(\Pi_\la(t) =  k) = \Pr(\NB( \alpha,\frac{\beta}{\beta + t}) = k), \ k=0,1,...
\ene
where $\NB( \alpha,p)$ denotes a random variable
which has a negative binomial distribution with size $\a$  and
probability $p$:
$$
\Pr(\NB( \alpha,p)= k) = \frac{\Gamma(\alpha + k)}{k!\ \Gamma(\alpha)}p^{\alpha}{(1 - p)}^{k},\ k = 0,1,..
$$
As in R programming language there are
standard functions for computing a negative binomial distribution,
relation (\ref{NB1}) can be used
for calculating distributions of PG processes.

For example, the distribution (\ref{PG}) can be calculated using a function in R:
\begin{verbatim}
dnbinom(k,size=alpha,prob=beta/(beta+t))
\end{verbatim}
To calculate CPF, $\Pr(\Pi_\la(t) \le L)$, where $\la=\Gam(\alpha,\beta)$, we can use a function
\begin{verbatim}
pnbinom(L,size=alpha,prob=beta/(beta+t))
\end{verbatim}

Now let us return to modeling enrollment.
Denote by $n_i(t)$ the enrollment process in site $i$ (the
number of patients recruited in  time interval $[0,t]$).

Denote also by $u_i$  the date of the activation of site $i$.
These dates at the initial stage may not be known in advance,
e.g., $u_i$ can be  considered as a uniform random variable in some
interval $[a_i,b_i]$ (\cite{an-dow-fed07,an08,an09a}).
The cases of beta and gamma distributions are considered in
\cite{an20}.
%(Anisimov, 2009a),
However, to avoid rather complicated calculations, we restrict
our attention to the case when the values  $u_i$ are known.

Then PG enrollment model assumes that in site $i$
the enrollment process $n_i(t)$ is
%is described as
a mixed Poisson process with
rate $\la_i$ in time interval $[u_i,\infty)$
where  $\lambda_i$
is viewed as a gamma distributed variable $\Gam(\alpha_i,\beta_i)$.
Thus, $n_i(t)$ is a Poisson process with time-dependent
rate $\lambda_i(t)$, where
$\lambda_i(t) = 0$ as $t \le u_i$ and $\lambda_i(t) = \lambda_i$ as $t > u_i$.

Consider also more convenient representation via a cumulative rate.
Denote $x(t,u) = max(0,t-u)$ (the duration of active enrollment at time $t$
for a site activated at time $u$).
So, if site $i$ is active at time $t$,
then $x(t,u_i) = t-u_i$.

Then the cumulative rate of the process $n_i(t)$
is $\La_i(t,u_i) = \la_i x(t,u_i)$.
This means,
if $\la_i=\Gam(\alpha_i,\beta_i)$,
$n_i(t)$ is a PG process with parameters $(x(t,u_i), \a_i, \be_i)$,
and the distribution of $n_i(t)$ can be calculated using (\ref{PG})
where in the right-hand side we should use $x(t,u_i)$ instead of $t$,
and parameters $(\alpha_i,\beta_i)$.

\subsection{Modeling enrollment on country level}\label{sec2-2}

Consider some country $s$ with $N_s$ sites. Denote by $I_s$ the set of indexes of these sites.
Then the enrollment process
$n(I_s,t)$ in this country is a mixed Poisson process with the
cumulative rate
\bee{CumRate}
\Lambda(I_s,t) = \sum_{i \in I_s}\la_i x(t,u_i).
\ene

Consider a special case when the rates in all sites
in this country have the same parameters $(\a,\be)$ of a gamma distribution.
Assume in addition that all $u_i \equiv u$ and $t > u$.
Then for all $i$, $x(t,u_i) = t-u $.
In this very special case in distribution
\bee{CumRate2}
\Lambda(I_s,t) = (t-u) \Gam(\a N_s, \be )
\ene

Thus, $n(I_s,t)$ is a PG process with parameters $(t-u,\a N_s, \be )$
and we can  use again relation (\ref{PG}) to calculate its distribution.

However, in practice we should not expect that all
sites will be activated at the same time.
Moreover, the parameters of the rates can be also different.
In these  cases, as the sum of gamma distributed variables with different
rate parameters $\be_i$ does not have a gamma distribution,
the cumulative rate $\Lambda(I_s,t)$
may not have a gamma distribution.
Therefore, the process $n(I_s,t)$ in general is not a PG process.

Thus, to develop the analytic technique for calculating the distribution of
$n(I_s,t)$ we have to use some approximations.

If $N_s$ is large enough ($N_s > 10$), in \cite{an11a} it was proposed a normal
approximation which used the closed-form expressions for the
mean and the variance of the rate $\Lambda(I_s,t)$.
This approximation works perfectly well for global enrollment.
However, for country predictions the normal approximation may not be appropriate
as in real trials in many countries the number of sites can be less than 10.
Therefore, for predicting
enrollment on country level we have to develop
another type of approximation that works efficiently for small
number of sites.

In \cite{an11a} it was proposed an approach to approximate
the country processes by PG processes
with some aggregated parameters
which was elaborated in details in \cite{an-aus20}.

Below we provide the details of this approach
as
it is essentially used in the paper for modeling the restricted enrollment
and for creating an optimal trial design.

At any time $t$ the cumulative rate of the enrollment process $n(I_s,t)$
is defined by (\ref{CumRate}).
Consider a general case where the rates $\la_i$ are gamma distributed with different parameters
$(\a_i,\be_i)$.
Denote for the ease of notation $v_i = x(t,u_i)$ - the duration of active enrollment in site $i$.
Then for a site active at time $t$, $v_i = t-u_i$, and clearly only these sites
can contribute to the number of patients enrolled up to time $t$.
Denote also by $m_i=\a_i/\be_i$ and $s_i^2=\a_i/\be_i^2$ the mean and the variance of $\la_i$
and introduce the mean and the variance of the cumulative rate
$\Lambda(I,t)$ as \
$
E(I_s,t) = \E [\Lambda( I_s,t )],\
S^{2}(I_s,t) = \Va[\Lambda( I_s,t )].
$
It is easy to see that
\bee{e20}
E(I_s,t ) = \sum_{i\in I_s} m_i v_i,  \
S^{2}(I_s,t) =  \sum_{i\in I_s} s_i^2 v_i^2
\ene

Let us introduce the variables
\bee{e21}
A(I_s,t) = E^{2}(I_s,t) / S^{2}(I_s,t), \
B(I_s,t) = E(I_s,t) / S^{2}(I_s,t)
\ene

The following statement is a slight extension of the result in \cite{an-aus20}
to the case where the rates are gamma distributed with different parameters.

%{\it \bf Lemma}\label{Lem1}
\belem{Lem1}
The distribution of $n(I_s,t)$
can be well approximated by
the distribution of a PG random variable
$PG(A(I_s,t),B(I_s,t))$.
\enlem

In \cite{an-aus20} it is shown using numerical calculations that
this approximation provides a very good fit even for a small number of
sites, 2,3, and with the larger number of
sites the difference between
the exact and approximative distributions is decreasing (see Appendix \ref{app1}).

The explanation of this result is the following. The cumulative rate
$\Lambda(I_s,t)$ has the same mean and the variance as
a gamma distributed  variable $\Gam(A(I_s,t),B(I_s,t))$.
Thus, the distribution of $n(I_s,t)$ can be approximated by the distribution
of the variable $\Pi(\Gam(A(I_s,t),B(I_s,t)))$ which by definition is $PG(A(I_s,t),B(I_s,t))$.

Note that this approximation resembles
in some sense
Welch-Satterthwaite \cite{satt46,welch47} approximation that was originally
used to approximate the linear combinations of independent chi-squared
random variables.

A PG approximation can be applied for any
number of sites and therefore is much more preferable compared to a
normal approximation, as provides a unified way for the approximation
of the global and country enrollment processes.

Using an approximation of the country process $n(I_s,t)$
by a PG process  $PG(A(I_s,t),$ $B(I_s,t))$,
we can calculate directly the mean value as \  $ \E [n(I_s,t)] = E(I_s,t)$
and, using formulae for a NB distribution, calculate the predictive bounds for any confidence
level $Q$.
Indeed,
$Q$-quantile of $n(I_s,t)$ can be calculated in R as
\begin{verbatim}
qnbinom(Q,size=A(Is,t),prob=B(Is,t)/(B(Is,t)+1))
\end{verbatim}
The quantiles for $Q=0.05$ and $Q=0.95$ reflect 90\%-predictive interval for $n(I_s,t)$.

It is also possible to calculate the distribution of the time to reach a specific target
for the number of patients in a country.
Denote by $\tau(I_s,L_s)$ the time to reach a given number of patients $L_s$ in country $s$.
As for any $t > 0$,
\bee{time-country}
\Pr( \tau(I_s,L_s) \le t ) = \Pr( n(I_s,t) \ge L_s)
\ene
the distribution of $\tau(I_s,L_s)$ is represented via
%the calculated
PG distribution of $n(I_s,t)$.

This provides a  useful opportunity to calculate the probabilities
to reach specific country goals and compare the performance of enrollment
in different countries.

\subsection{Modeling global enrollment}\label{sec2-3}

Assume that trial involves $S$ countries.
The global enrollment process $n(t)$ is a sum of country processes and
is a mixed Poisson process with the global cumulative rate
\bee{CumRateGlob}
\Lambda(t) = \sum_{s=1}^S \Lambda(I_s,t)
\ene
where country rates $\Lambda(I_s,t)$ are defined in (\ref{CumRate}).

Assuming for simplicity that all sites are active at time $t$ and
using relations (\ref{e20}) we get the relations for the mean $E(t)$
and the variance $S^2(t)$ of the rate $\Lambda(t)$,
\bee{MeanGlob}
E(t) = \sum_{s=1}^S E(I_s,t ); \ S^2(t) = \sum_{s=1}^S S^2(I_s,t ).
\ene
Then, using Lemma \ref{Lem1}, we can approximate the distribution of $n(t)$
by the distribution of a PG random variable
$PG(A(t),B(t))$, where
\bee{A-B-Glob}
A(t) = E^{2}(t) / S^{2}(t), \
B(t) = E(t) / S^{2}(t).
\ene

Using this approximation and formulae for a negative binomial distribution we can calculate
the mean, median and $Q$-predictive bounds
for the process $n(t)$.

Correspondingly,
denote by $\tau(n)$ the time to reach the planned number of patients $n$ (to complete enrollment).
As
\bee{time-global}
\Pr( \tau(n) \le t ) = \Pr( n(t) \ge n),
\ene
the probability to complete enrollment before time $t$
is represented via the calculated PG distribution of $n(t)$.
Therefore, PoS (to complete enrollment before a planned date $T_{plan}$)
is calculated as
\bee{PoS}
\Pr( \tau(n) \le T_{plan} ) = 1-\Pr( n(T_{plan}) \le  n-1).
\ene

\section{Modeling enrollment with restrictions}\label{sec3}

In this section we develop a novel technique for modeling
and forecasting enrollment under the upper restrictions (caps)
on country level.

\subsection{Modeling enrollment with restrictions in one site}\label{sec3-2}

Consider first modeling a restricted enrollment in one site.
Consider a site $i$ with the enrollment rate $\la_i =
\Gam(\a_i,\be_i)$ and time of activation $u_i$. Assume that the
enrollment in this site is stopped when the number of patients hits a given
upper threshold (cap) $L_i$. For the ease of notation, omit index $i$ at the
variables $\a_i,\be_i.u_i,L_i.$

Consider first the unrestricted process $n_i(t)$ and
denote by $P(k,t,u)$ its distribution
which is defined according to (\ref{PG}) as
\bee{PG-3}
P(k,t,u)
= \frac{\Gamma(\alpha + k)}{k!\ \Gamma(\alpha)}\ \frac{x^k(t,u)\beta^{\alpha}}
{(\beta + x(t,u))^{\alpha + k}}, \ k = 0,1,..
\ene

Define now the enrollment process  $n_i^{L}(t) $ restricted by cap $L$ as
\begin{equation}\label{e1}
n_i^{L}(t) =
\left \{
 \begin{array}{lcc}
n_i(t) & as  &  n_i(t) < L \\
L & as  &  n_i(t) \ge L \\
 \end{array}
 \right.
        \end{equation}

Then the distribution of $n_i^{L}(t)$ can be calculated directly:
\begin{equation}\label{e2}
\Pr(n_i^{L}(t) = k) =
\left \{
 \begin{array}{lcc}
P(k,t,u) & as  & 0\le k < L \\
1- \sum_{k=0}^{L-1} P(k, t,u)  & as  &  k = L \\
0 & & \hbox{otherwise}
 \end{array}
 \right.
        \end{equation}

Correspondingly, the first two moments are calculated as follows (see Section \ref{App-mean}
and Section \ref{App-2nd}
in Appendix):
\beeq{Ecap2-1}
\E[n_i^{L}(t)]  &=& \frac {\a x(t,u)} {\be} \Pr(\PG(x(t,u), \a+1,\be)) \le L-2)) \nn \\
&+&
L \Big(1- \Pr(\PG(x(t,u), \a, \be) \le L-1)\Big)
\eneq

\beeq{Cap2-1}
\E[(n_i^{L}(t))^2]
 &=&  \frac {\a(\a+1) x^2(t,u) }{\be^2} \Pr(PG(x(t,u),\a+2,\be)\le L-3) \nn \\
&+& \frac {\a x(t,u)}{\be} \Pr(PG(x(t,u),\a+1,\be)\le L-2) \\
 &+&
L^2 \Big(1- \Pr(PG(x(t,u),\a,\be) \le L-1)\Big) \nn
\eneq

\subsection{Modeling enrollment with restrictions on country level}\label{sec3-3}

In real trials typically restrictions can be imposed on country level
based on some regulatory assumptions.
Using the results of Sections \ref{sec2-2} and \ref{sec3-2}
we can develop an analytic technique for predicting  restricted
enrollment on country level.

Consider some country $s$ with $N_s$ sites indexed by set $I_s$.
According to Lemma \ref{Lem1}, the distribution of unrestricted enrollment process
$n(I_s,t)$ in this country can be well
approximated by the distribution
of a PG variable  $\PG(A(I_s,t),B(I_s,t))$ which has the same distribution
as a PG process $\PG(1,A(I_s,t),B(I_s,t))$.
That means, for the distribution of $n(I_s,t)$ we can use formula (\ref{PG-3}) where in
the right-hand side we should put
$x(t,u)=1, \a = A(I_s,t), \be = B(I_s,t)$.

Assume now that there is a cap $L(s)$, so the
enrollment in country $s$ is stopped when the number of patients
$n(I_s,t)$ reaches
$L(s)$. To model the process $n(I_s,t)$ restricted by
cap $L(s)$, (denote it as $n^{L(s)}(I_s,t)$) we can use the same
relations as in Section \ref{sec3-2} above, where
we should put $x(t,u)=1, \a = A(I_s,t), \be = B(I_s,t)$.

Then, according to (\ref{e2}),
the distribution of a restricted process in country $s$ is defined as:
\beeq{Count1}
\Pr( n^{L(s)}(I_s,t) = k ) &=& \frac{\Gamma(A(I_s,t) + k)}{k!\ \Gamma(A(I_s,t))}\
\frac{B(I_s,t)^{A(I_s,t)}} {(B(I_s,t) + 1)^{A(I_s,t) + k}}, \ k = 0,1,..,L(s)-1,  \nn \\
\Pr( n^{L(s)}(I_s,t) = L(s) ) &=& 1- \Pr(\PG(A(I_s,t), B(I_s,t)) \le L(s)-1)
\eneq

Correspondingly, using relations (\ref{e21}), (\ref{Ecap2-1})
and (\ref{Cap2-1}) we get
\beeq{Ecap2-2}
\E[n^{L(s)}(I_s,t)]  &=& E(I_s,t) \Pr(\PG(A(I_s,t)+1,B(I_s,t)) \le L(s)-2)) \\
&+&
L(s) \Big(1- \Pr(\PG(A(I_s,t), B(I_s,t)) \le L(s)-1)\Big) \nn
\eneq

\beeq{Cap2-2}
\E[(n^{L(s)}(I_s,t))^2]
 &=&  (E^2(I_s,t)+S^2(I_s,t)) \Pr(PG(A(I_s,t)+2,B(I_s,t)))\le L(s)-3) \nn \\
&+& E(I_s,t) \Pr(PG(A(I_s,t)+1,B(I_s,t))\le L(s)-2) \\
 &+&
L^2(s) \Big(1- \Pr(PG(A(I_s,t),B(I_s,t)) \le L(s)-1)\Big) \nn
\eneq

Then
\beeq{VarCap}
\Va[n^{L(s)}(I_s,t)] = \E[(n^{L(s)}(I_s,t))^2] - \Big( \E[n^{L(s)}(I_s,t)] \Big)^2
\eneq

Consider an important characteristic -- the time $\tau(I_s,L(s))$ to reach
cap $L(s)$ in country $s$.
According to Lemma \ref{Lem1}, we can use the following relation:
for any $t > 0$,
\beeq{time-country-2}
\Pr( \tau(I_s,L(s)) \le t ) &=& \Pr( n(I_s,t) \ge L(s)) \nn \\
&=& 1 - \Pr( PG(A(I_s,t),B(I_s,t)) \le L(s)-1).
\eneq

\subsubsection{Asymptotic properties}\label{sec3-4}

Consider the asymptotic dependence of the country enrollment process restricted by
cap $L(s)$ on the time
and on the number of sites.

{\bf 1st case}. Consider the case where $t \to \infty$.
Denote
$$
M(I_s) = \sum_{i\in I_s} m_i, \ V^{2}(I_s) =  \sum_{i\in I_s} s_i^2.
$$

%\belem{Lem2}
%{\it Lemma}\label{Lem2}
%\begin{lemma}\label{Lem2}
\belem{Lem2}
Assume that $t \to \infty$ and other parameters are fixed.
Let also $M(I_s) > 0$, $V^{2}(I_s) >0$.
Then,
\bee{Asymp1}
n^{L(s)}(I_s,t) \pto L(s)
\ene
where symbol $\pto$ means convergence in probability.
%\enlem
\end{lemma}

{\em Proof.}
As \ $t \to \infty$, in relation (\ref{e20}),
$$
E(I_s,t) = M(I_s)t (1+O(1));  \ S^2(I_s,t) = V^{2}(I_s) t^2 (1+O(1))
$$
Thus, in relation (\ref{e21}),
$$
A(I_s,t) \to M^2(I_s)/V^{2}(I_s) >0, \ B(I_s,t) = O(1/t).
$$
From relation (\ref{PG}), for any $k \ge 0, \a > 0$, as $t \to \infty$
and $\be = O(1/t)$,
$$
\Pr(\PG(\alpha,\beta) =  k) \to 0; \
t \Pr(\PG(\alpha+1,\beta) =  k) \to 0; \
t^2 \Pr(\PG(\alpha+2,\beta) =  k) \to 0.
$$

Using these relations together with (\ref{Count1})
we get from  (\ref{Ecap2-2}), (\ref{Cap2-2}):
\bee{Asym1}
\E[n^{L(s)}(I_s,t)] \to L(s); \ \E[(n^{L(s)}(I_s,t))^2] \to L^2(s).
\ene
Thus, \
$
\Va[n^{L(s)}(I_s,t)] \to 0,
$
and relation (\ref{Asymp1}) follows from Chebyshev's inequality.
\\

Actually, for a restricted process
the relation
(\ref{Asymp1}) is expected.

Note that the case  $V^{2}(I_s) = 0$ corresponds to a Poisson
model with fixed rates and can be considered similarly.

{\bf 2nd case}.
Consider now the case where the number of sites $N_s \to \infty$.

%E(I_s,t) = M(I_s)t (1+O(1));  \ S^2(I_s,t) = V^{2}(I_s) t^2 (1+O(1))

%{\it Lemma}\label{Lem2-2}
\belem{Lem2-2}
Assume that for any $t > 0$, $E(I_s,t)/N_s \to \wti M_s(t)$, \
$S^2(I_s,t)/N_s \to \wti V^2_s(t)$,
where $\wti M_s(t)$ and $\wti V^2_s(t)$ are some bounded functions,
and $\wti M_s(t) >0$, $\wti V^2_s(t) > 0$.

Then relation (\ref{Asymp1}) holds.
\enlem

{\em Proof.}
In this case, \
$
A(I_s,t) = O(N_s) \to \infty, \ B(I_s,t) \to \wti M_s(t)/\wti V^2_s(t) > 0.
$

Note that as $\a \to \infty$, in relation (\ref{PG}), for any $k \ge 0$,\
$\Gamma(\alpha + k)/\Gamma(\alpha) = O(\a^k)$,
and for any $q$, $0 < q < 1$, $\a^k q^\a \to 0.$
Thus, for any $p > 0$, $k \ge 0$,
$$
\a^p \Pr(\PG(\alpha,\beta) =  k) \to 0.
$$
Therefore, similar to Case 1,
using these relations together with (\ref{Count1})
we get
the relation (\ref{Asym1}).
Finally, relation (\ref{Asymp1}) follows from Chebyshev inequality.
\\

This result shows that for rather large number of sites in a
country, the country cap can be reached rather quickly, earlier than
the planned stopping time, and after that point this country will not
contribute further into the global enrollment. Thus, the caps should
be chosen rather carefully by analyzing and comparing the times to
reach country caps with the planned enrollment time.

For example, denote by $T$ the planned enrollment time and assume
that in relation (\ref{time-country-2}), $\Pr( \tau(I_s,L(s)) \le T )$
is rather high (say, more than 0.9). Then it is very likely that
the cap in this country will be reached before the planned time $T$.
Thus, if the enrollment will go according to plan,
the sites in this country will not be used fully
efficiently.
If there are many caps in different countries such that
these caps can be reached with high probabilities before time $T$,
this will lead to closing of enrollment in these countries
earlier than planned which may lead to substantial delay
of the global enrollment.

Therefore, in these cases it can be recommended to reconsider the design
of enrollment and increase or eliminate caps in such countries if possible.

\subsection{Forecasting global enrollment under country restrictions}\label{GlobRest}

Consider now forecasting of the global enrollment when there are
enrollment caps in some countries.

The global enrollment process
is a sum of restricted by caps $L(s)$ country PG processes $n^{L(s)}(I_s,t)$.
As in every country the distribution of $n^{L(s)}(I_s,t)$ is explicitly defined by (\ref{Count1}),
then the distribution of the global process can be calculated numerically
using a convolution of the country processes.
In R-software this distribution can be calculated using very fast numeric procedure based
on a discrete Fourier transform and R-function {\em convolve()}.
Let's call this approach a "distributional approach".

This algorithm is working very efficiently and calculates for any $t$
the vector distribution of the global process.
Using this distribution we can calculate numerically
the predictive mean, median and predictive bounds.

Correspondingly, at any time $t$, using relation (\ref{time-global})
for the global enrollment time $\tau(n)$
and the distribution of the enrollment process $n(t)$ calculated using a convolution of country
processes,
we can also calculate the probability to complete enrollment before time $t$
and the probability to complete
before the planned time $T_{plan}$ which is a PoS.

Note that when the
number of countries is rather large
(more than $10$), for the global enrollment process $n(t)$
we can also use a normal approximation.
Indeed, using expressions
(\ref{Ecap2-2}) and (\ref{VarCap}) for the mean and the variance of country processes,
for any $t > 0$ we can calculate the mean and the variance
of the global enrollment process $n(t)$ (as sums of means and variances
of country processes)
and use them to calculate the predictive bounds and PoS
based on a normal approximation similar as it was
considered for unrestricted process in Section \ref{sec2-3}
(see also \cite{an11a}).

Correspondingly, at any time $t$, using relation (\ref{time-global})
for the global enrollment time $\tau(n)$  and the approximative normal
distribution of the enrollment process $n(t)$,
we can also calculate the probability to complete enrollment before time $t$
and PoS.

Note that calculations based on using a normal approximation take much less computational time
compared to a "distributional approach".
However, to evaluate one particular scenario, it takes nearly invisible time by using
any approach.
Therefore, for the purpose of creating predictions for several particular scenarios,
it can be recommended using a "distributional approach" which is a universal approach
as can be applied for any number of the countries.

Another situation is when we consider an optimal design and need to run a huge
number of different scenarios. This case will be discussed later
in Section \ref{Opt-Restr}.

Note that for practical reasons it is enough to provide calculations on a daily basis.
Therefore, to create the predictions of country and global enrollment processes,
we need first to evaluate the upper predictive bound for the enrollment time
using rather high confidence level (usually 0.95).
This can be done numerically using (\ref{time-global}) and calculating
sequentially the first
time $T_{0.95}$ such that
\bee{time-bound}
\Pr( n(T_{0.95}) \ge n ) \ge 0.95.
\ene
Then we consider a sequence of times $t_k$
(usually $(1,2,..,T_{0.95})$),
and for every $t_k$ calculate numerically the predictive mean, median and the bounds for $n(t_k)$
for a given confidence level (usually 0.9) using the calculated distribution of $n(t_k)$.
Probability to complete enrollment up to any time $t_k$ can be calculated using (\ref{time-global}).

%The main probability of interest (Probability of Success - PoS) is the probability to complete
%enrollment before the planned time $T_{plan}$
%which is  $\Pr( \tau(n) \le T_{plan} ) $. Then PoS can be calculated using
%the distribution of $n(t)$ at time
%$t = T_{plan}$ and (\ref{time-global}).

Note that PoS plays an important role at the initial study design.
If PoS is not very large, then
it is likely
that study can be delayed. Therefore, it can be recommended to improve the enrollment design
where one of the options can be adding more clinical sites and recalculating  PoS.

\subsubsection{Analysis of the impact of enrollment caps}\label{sec3-6}

In Section \ref{sec3-3} it is noted that the enrollment caps in countries may lead to a substantial delay
of the global enrollment
and to the inefficient use of sites in these countries.
Consider some numeric approaches for the analysis and comparing the impact of caps.

Assume that there are several countries $(1,..,J)$ with restrictive caps $L(j)$.
Using relation (\ref{time-country-2}) and formula for the distribution of the
unrestricted PG process $n(I_j,t)$ in country $j$, the probability $P(T,I_j,L(j))$
to reach cap in this country before the planned enrollment time $T$ is calculated as
$$
P(T,I_j,L(j)) =  1 - \Pr( n(I_j,T) \le L(j)-1)
$$
Correspondingly, using the results of Section \ref{GlobRest}
we can calculate the PoS $P(T)$ to complete the global enrollment before time $T$.

Now, if for country $j$, $P(T,I_j,L(j)) > P(T)$, then it is likely that the
cap $L(j)$ in this country will be reached before stopping the global enrollment.
Thus, for country $j$ it can be recommended to increase the value of cap if possible.

Another opportunity is to compare the quantiles of the times to reach caps with the quantile
of the global enrollment time.

Consider some value $Q$ (e.g. $Q=0.9$).
Using formula (\ref{time-country-2}) for the distribution of the time
$\tau(I_j,L(j))$,
we can
calculate its $Q$-quantile $S(Q,j,L(j))$.

Now, using the results of Section \ref{GlobRest}, we can calculate $Q$-quantile $S(Q,n)$
of the global enrollment time $\tau(n)$.
Then we can compare the values $S(Q,j,L(j))$ and $S(Q,n)$.
If for some country $j$, $S(Q,j,L(j)) < S(Q,n)$, then it is likely that the
cap in country $j$ will be reached before stopping the global enrollment.
Thus, for this country it can be recommended to increase the value of cap if possible.
It can be also proposed to compare the mean times in countries to reach caps and the mean
of the global enrollment time, however, this approach in general leads to similar conclusions.

\begin{figure}\label{Fig1}
\centering
\includegraphics[width=14.0cm,height=7cm]{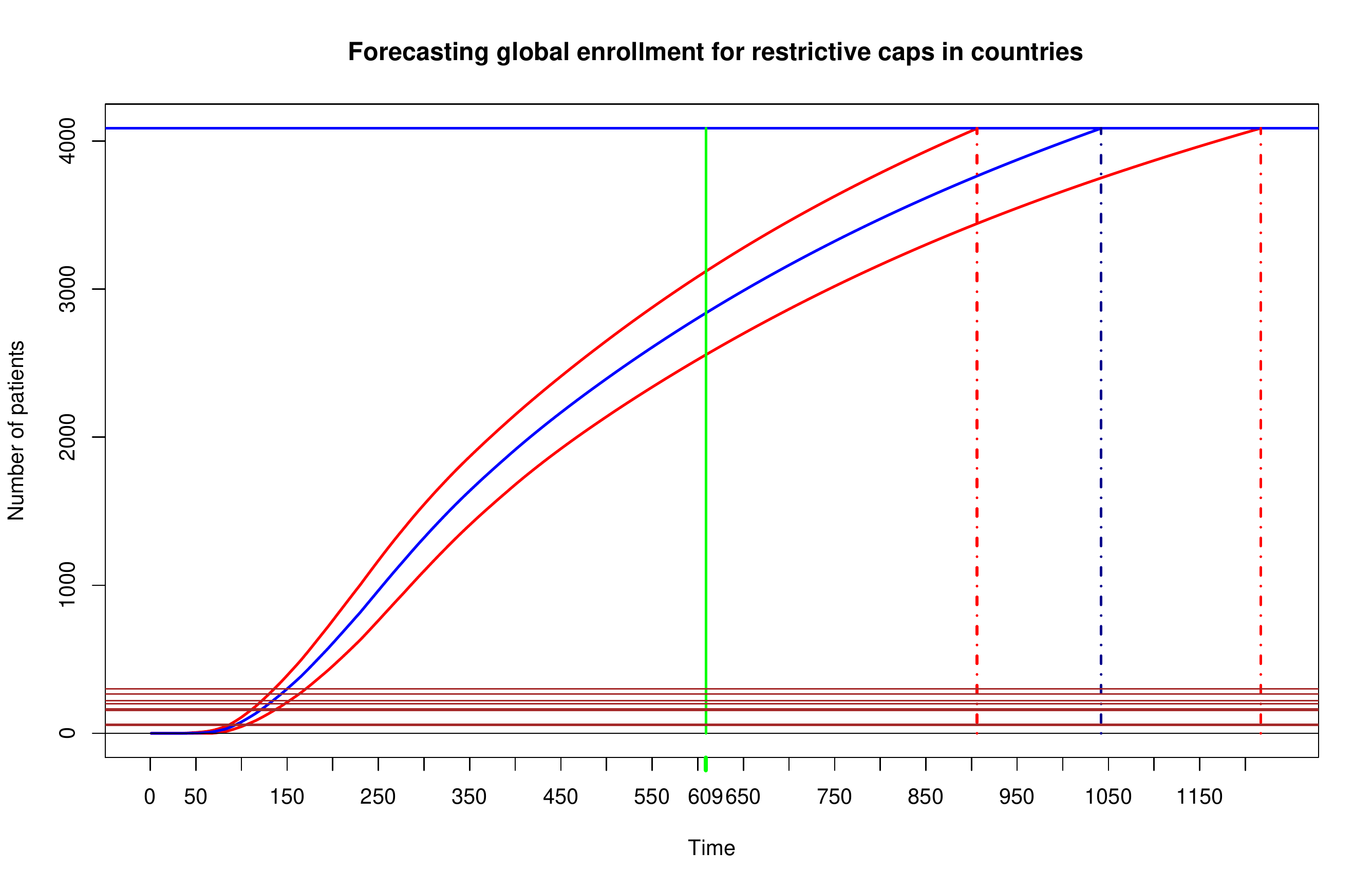}
\caption{ \small
Forecasting the enrollment with mean and 90\% predictive bounds for some initial set-up of country caps.
 The values of caps in each country are set voluntary
and are shown by brown horizontal lines. These are the caps in the first 10 countries:
(265, 200, 200, 200, 200, 200, 200, 200, 200, 55).
For this case, the predictive mean and 90\% predictive interval to complete enrollment in days are (1044, 906, 1217).
Probability to complete in time (PoS) is zero.
}
\end{figure}
\begin{figure}\label{Fig2}
\centering
\includegraphics[width=14.0cm,height=7cm]{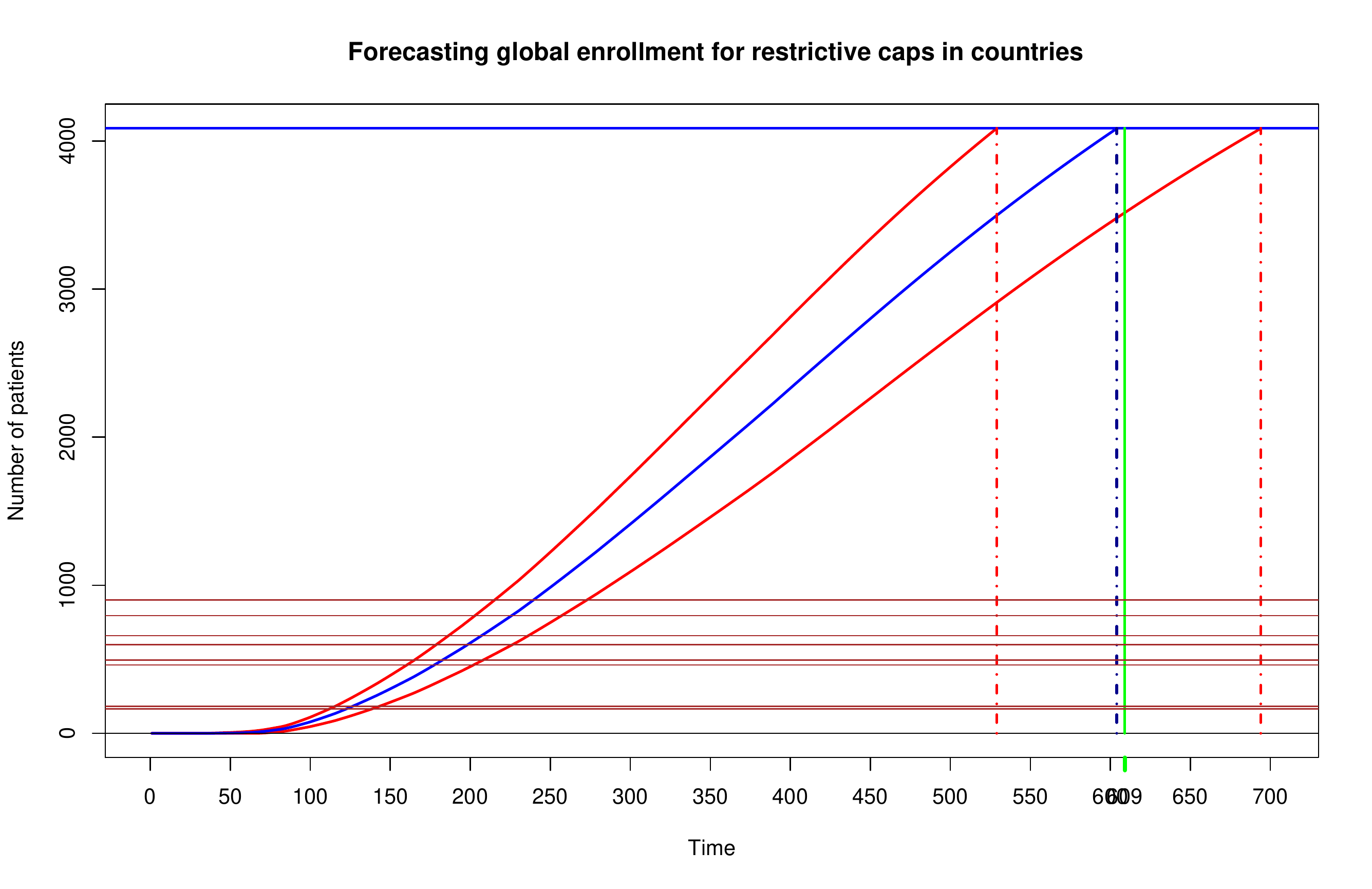}
\caption{ \small
Forecasting the enrollment with mean and 90\% predictive bounds where the initial caps are increased on 200\%.
The values of caps in each country are shown by brown horizontal lines.
These are the caps in the first 10 countries:
(795, 600, 600, 600, 600, 600, 600, 600, 600, 165).
For this case, the predictive mean and 90\% predictive interval to complete enrollment in days are (606, 529, 694),
about 14 months earlier compared to the previous design.
PoS is 0.537.
}
\end{figure}

Consider as a case example a realistic study with 4100 patients planned to be recruited.
The planned enrollment time is 20 months.
There are 40 countries and it is set some start-up enrollment design
(for every country - the number of sites to be initiated, sites initiation dates, and
the mean and the coefficient of variation of the enrollment rates).

Figures 1 and 2 show a potential impact of country caps on the global enrollment time
which can be substantial for not so large caps.

As it is seen, for this study with the initial setup of caps, the increase in the mean enrollment time
is about 14 months
and PoS is zero. Thus, the study cannot be completed in time before 20 months.
But with the increased caps, the study on average is going as planned.

Therefore, the enrollment design involving country restrictions should be first evaluated
by analyzing the impact of caps in different countries on the whole duration of the enrollment,
as for not so large caps the increase in the enrollment time can be dramatic.

\subsection{Using historic data for better prediction of the enrollment rates
for the new trials}\label{historic}

The technique for modeling and forecasting enrollment
uses  some input parameters, specifically, the mean and the variance of the enrollment rates
and the times of sites' activation.

Actually at the initial (planning) stages the enrollment rates are not known in advance.
Therefore, a practical question of a paramount interest is:
how to estimate efficiently the parameters of
the rates at the planning stage when real trial data are not available yet using historic data
from similar trials?

Typically, the enrollment rates are provided by clinical teams using the expert estimates
and their knowledge of the specifics of particular trials.

At the current stage pharmaceutical companies have an access to very large databases of historic trials.
These data can be used to evaluate the values of historic rates and these can be used
as the initial rates for the new trials.

As there are many factors which can influence the enrollment, standard regression models
may not work well. Therefore, one of the directions is using machine learning algorithms
trained on large databases of historic studies using different features:
therapeutic area, study indication, number of sites, study start-up times,
phase, country, enrollment windows, etc.

This is a very important area which requires a separate discussion.
Some approaches on using a PG model for predicting new trials were proposed
in \cite{savy17}.

\section{Optimal enrollment design}\label{optimal}

One of the cornerstone problems at the planning stage is: find an optimal allocation of
sites/countries that minimizes the total trial cost
given that PoS
is no less than a given value
and there are certain restrictions on the number of sites
in countries.

To formalize this problem let us introduce the basic notation.

Consider a given set of countries $(1,..,S)$ and
assume that we have chosen some number of sites $(N_1,..,N_S)$ in these countries.
Let $T = T_{plan}$ is the target enrollment time.
Suppose also that for any given country $s$ and the value $N_s$,
the times of sites' activation $(u_1^{s},..,u_{N_s}^{s})$ are generated according to some
algorithm, e.g. it can be a uniform grid on some interval $[a_s,b_s]$
or piece-wise uniform grid using the expected quartiles of the times of sites' activation
(e.g. the times when 25\%, 50\%, 75\%, 100\% of sites to be activated).

Assume for simplicity that the mean and the variance
$(m(s),\si^2(s))$
of the enrollment rates in any country $s$ are the same for all sites in this country
and all sites are planned to be activated before target time $T$.

Consider the following costs:

\begin{enumerate}

\item
the vector of costs per selecting one site in each country, $\bar C = (C_s, s=1,..,S) $;

\item
the vector of costs per one enrolled patient in each country, $\bar c = (c_s, s=1,..,S)$;

\item
the vector of costs per including country $s$ with non-zero number of sites,
$\bar Q = (Q_s, s=1,..,S) $;

\end{enumerate}

Denote by $C(T, \bar N, \bar C, \bar c, \bar Q)$ the total mean cost of the trial
in time interval $[0,T]$ for a given site's allocation $\bar N = (N_1,..,N_S)$.

Assume also that there is some planned set of restrictions $W$ on the number of sites, e.g.
the  minimal and maximal number of sites for each country.

Denote by $P(n,T,\bar N)$ a PoS -- the probability to reach a planned number of patients $n$
for a given site's allocation $\bar N = (N_1,..,N_S)$ before target time $T$.

Then the optimal enrollment design is a solution of the following
problem: \\

{\em Optimization problem 1:}

For a given probability $P$ find an optimal site's allocation $\bar N = (N_1,..,N_S)$ that \\
minimizes the total cost \ $C(T, \bar N, \bar C, \bar c, \bar Q)$ \
given
\bee{PoS-0}
P(n,T,\bar N) \ge P
\ene
$$
\bar N \in W
$$
where $P$ is an agreed confidence level (e.g. 0.8, 0.9,..).

\subsection{Unrestricted enrollment}

Consider two main approaches in the case of unrestricted enrollment for how to calculate
the PoS and the optimal trial design
 depending on whether the number of countries $S$ is rather large or not.

\subsubsection{The number of countries is rather large}

Assume that $S > 10$, so we can use a normal approximation
for the global enrollment process $n(T) = n(T, \bar N)$ as a sum of country
processes $n(I_s,T)$.

The global cumulative enrollment rate at time $T$ has the form
\bee{CumRate3}
\La(T, \bar N) = \sum_{s=1}^S \sum_{i \in I_s}
\la_i^{s} (T - u_i^{s}),
\ene
where $\la_i^{s}$ are the enrollment
rates in sites in country $s$ with mean $m(s)$ and variance
$\si^2(s)$ and we assume for simplicity that all sites are initiated before time $T$.
Therefore, values $E(I_s,T)$ and $S^{2}(I_s,T)$ defined in (\ref{e20}) have the form:
\bee{e20-2}
E(I_s,T) = m(s) \sum_{i\in I_s} (T- u_i^{s}),  \
S^{2}(I_s,T) =  \si^2(s) \sum_{i\in I_s}(T- u_i^{s})^2,
\ene
and the mean $E(T,\bar N)$ and the variance $S^2(T,\bar N)$ of
$\La(T, \bar N)$
are expressed as
\bee{CumRate-mean}
E(T,\bar N) = \sum_{s=1}^S E(I_s,T), \
S^2(T,\bar N) = \sum_{s=1}^S S^{2}(I_s,T).
\ene

Denote
\bee{G2}
G^2(T,\bar N) = E(T,\bar N) + S^2(T,\bar N).
\ene
Note that $G^2(T,\bar N) = \Va[n(T)]$.
Using relation (\ref{time-global}) and a normal approximation for the process $n(T)$
we can easy derive the following criterion:
\\

{\em Criterion (to complete enrollment in time):}

The study for a chosen country's allocation $\bar N$
will complete enrollment up to time $T$ with probability $P$
if the following inequality is satisfied:
\bee{Crit1}
\frac {E(T,\bar N) - n}{\sqrt{G^2(T,\bar N)}} \ge z_{P}
\ene
where $z_{P}$ is a $P$-quantile of a standard normal distribution.

Consider now the calculation of global costs.
The cost for sites involved is
\bee{cost1}
Cost(sites, \bar N) = \sum_{s=1}^S C_s N_s
\ene
The cost for the mean number of patients recruited in interval $[0,T]$ is
\bee{cost2}
Cost(patients, \bar N) = \sum_{s=1}^S c_s m(s) \sum_{i \in I_s} (T - u_i^s)
\ene
The cost for the countries with non-zero number of sites is
\bee{cost3}
Cost(countries, \bar N) = \sum_{s=1}^S Q_s I(N_s > 0)
\ene
where $I(A)$ is the indicator of the event $A$.

Thus, for any given allocation of sites $\bar N$,
the global cost $C(T, \bar N, \bar C, \bar c, \bar Q) $
is the sum of costs defined by relations (\ref{cost1})-(\ref{cost3}).

Note also that the condition $\bar N \in W $ typically has the following form:

define the vector $\bar H = (H_1,..,H_S)$ of the low bounds and the vector $\bar U = (U_1,..,U_S)$
of the upper bounds for the number of sites in each country. Then the condition
$\bar N \in W $ means:
\bee{sites-bounds}
H_s \le N_s \le U_s, \ s = 1,2,.., S.
\ene

In this setting, the optimization problem has the following general form:
\\

{\em Optimization problem 2:}

For a given probability $P$ find an optimal site's allocation $\bar N$ that:

minimizes the global cost $C(T, \bar N, \bar C, \bar c, \bar Q) $
given conditions (\ref{Crit1}) and (\ref{sites-bounds}).
\\

Note that the set of possible allocations should not be empty, so the probability $P$
can be reached for some allocation. This will be guaranteed if the following
condition is satisfied:

{\em Condition of feasibility for probability $P$:}
\bee{Feasible}
\frac {E(T,\bar U) - n}{\sqrt{G^2(T,\bar U)}} \ge z_{P}
\ene

As the total cost and condition (\ref{Crit1})
have a non-linear dependence on vector $\bar N$, this general problem can be solved
using the methods of constrained optimization or random search.

\subsubsection{Approach using step-wise linearisation}\label{linear}

Assume that in restrictions (\ref{sites-bounds}) for all $s$, $H_s > 0$,
so all countries at the design stage
plan to involve some sites, which is quite natural.
Assume for simplicity that
the times $(u_1^{s},..,u_{N_s}^{s})$ of sites' activation
in country $s$ are chosen as a uniform grid in some time interval $[a(s),b(s)]$ defined
for this country at the planning stage. In general it can be considered
more sophisticated algorithms.

For a given sites' allocation $\bar N = (N_1,..,N_S)$, define
for every country $s$, assuming $N_s > 0$,
the average enrollment time $R(s)$ for any generic site in this country:
\bee{mean-time-site}
R(s) = \frac 1 {N_s} \sum_{i \in I_s} (T - u_i^s)
\ene

In this case
\bee{ETN}
E(T,\bar N) = \sum_{s=1}^S N_s m(s) R(s)
\ene
and the patient cost in (\ref{cost2}) can be written as
\bee{cost2-2}
Cost(patients, \bar N) = \sum_{s=1}^S N_s c_s m(s)  R(s).
\ene

Note that in the case when in country $s$ the values $u_i^s$ are generated according to a
uniform distribution in interval $[a(s),b(s)]$, the mean country enrollment time is
\bee{mean-time-site-2}
R(s) = \frac 1 {N_s} \E [\sum_{i \in I_s} (T - u_i^s)] = \E (T - u_1^s) = (b(s)-a(s))/2,
\ene
so $R(s)$ doesn't depend on $N_s$.
Thus, we can keep a linear representation (\ref{cost2-2}) for any other vector
of the number of sites
in countries assuming
that the times of activation of sites in country $s$
are chosen as a uniform grid.

Using this representation, we see that all costs are linearly dependent
on the running vector of sites $\bar N$.
This representation essentially accelerates the computations
on each step in the optimization algorithm.
At the final stage, when we will calculate the optimal number of sites,
we can exactly calculate PoS using
a specific sites' allocation in each country.
However, numerical calculations show that the difference
in PoS, calculated using a proportional method as above or
the specific uniform grid of sites' allocation,
is in the 2nd-3rd digit after comma. Thus, this approach
can be efficiently used in practice.

Now the remaining point is -- how to deal with a non-linear condition (\ref{Crit1}).
This condition can be written in the form
\bee{Crit1-2}
E(T,\bar N) - z_{P} \sqrt{G^2(T,\bar N)} \ge  n.
\ene
The value $E(T,\bar N)$ can be represented in a linear form with respect to vector $\bar N$
as in (\ref{ETN}).
The value $G^2(T,\bar N)$ in (\ref{G2}) can be also represented
in a linear form with respect to vector $\bar N$ using for every country an averaged quadratic
enrollment time in any generic site:
\bee{Vs}
V(s) = \frac 1 {N_s} \sum_{i \in I_s} (T - u_i^s)^2
\ene
Then according to (\ref{e20-2}),
$$
G^2(T,\bar N) = \sum_{s=1}^S N_s \Big( m(s) R(s) + \si^2(s) V(s) \Big)
$$
However, relation (\ref{Crit1-2}) is still non-linear with respect
to vector $\bar N$ except the case when $P=0.5$
as $z_{0.5}=0$.

To resolve this problem, it is developed a step-wise recurrent algorithm where on each
step we set linear restrictions
and use a simplex method for linear constrained optimization which is working
extremely fast even for very large
number of countries up to several hundreds.

Note that the simplex method assumes that the variables involved into optimization can take also
non-integer values.
Assuming so, we can find a solution of optimization problem in the space
of continuous variables, and then at the last
step, we can use a simple search checking for every non-integer
variable $x_k$ which of the two
nearest integer values, lower $N_k^{Low}$ or upper $N_k^{Upp}$,
gives the least total cost keeping
condition (\ref{Crit1-2}). On this way, we will find
a quasi-optimal discrete allocation
$\bar N_{opt}$ satisfying the conditions of optimization problem 2.

The description of the step-wise recurrent algorithm is the following.
First, for any running site's allocation $ \bar N$ we introduce
the new vector variable $\bar x = \bar N - \bar H$.

Then $E(T,\bar N) = E(T,\bar x) + E(T,\bar H)$ and
the global costs have the form:
$$
C(T,\bar N, \bar C, \bar c, \bar Q) =
C(T,\bar H, \bar C, \bar c, \bar Q) + C(T,\bar x, \bar C, \bar c, \bar Q),
$$
where $C(T,\bar x, \bar C, \bar c, \bar Q)$ depends linearly on $\bar x$, and \
$
0 \le \bar x \le \bar U - \bar H
$ \
by all components.

Now let us start with the initial vector $\bar x^{(0)} = \bar 0$ and find
the next value $\bar x^{(1)}$
as a solution of the optimization problem with linear constrains using
simplex method with respect
to vector $\bar x = (x_1,..,x_S)$,
where condition (\ref{Crit1-2}) is re-written to have linear restrictions on vector $\bar x$:
\bee{Crit1-3}
\sum_{s=1}^S x_s m(s) R(s)
\ge  n +  z_{P} \sqrt{G^2(T,\bar x^{(0)} + \bar H)} - E(T,\bar H)
\ene
Correspondingly, denote by $\bar x^{(k)}$ a solution of the linear constrained
optimization problem on step $k$.
The next value $\bar x^{(k+1)}$ is calculated as a solution of the
linear constrained optimization problem with respect
to vector $\bar x$
where
(\ref{Crit1-3}) has the form
\bee{Crit1-4}
\sum_{s=1}^S x_s m(s) R(s)
\ge  n +  z_{P} \sqrt{G^2(T,\bar x^{(k)} + \bar H)} - E(T,\bar H)
\ene

Convergence of this algorithm can be proved in one dimensional case.
Indeed, consider a trial with country $\{ 1 \}$ only. Assume for simplicity that $H_1=0$.
Then the relation (\ref{Crit1-4})
on step $k$ will be reduced to the relation
\bee{Crit1-5}
x^{(k+1)} =  \frac n E + \frac V E \sqrt{x^{(k)}}
\ene
where $x^{(0)} = 0$
and $x^{(k)} \le U_1$, and according to (\ref{e20-2})-(\ref{G2}), $E$ and $V$ are some constants,
specifically,
$E = m(1) T/2$, $V = z_{P} \sqrt{m(1)T/2 + \si^2(1)T^2/4}$.

Thus,
\beeq{recur}
x^{(1)} &=& \frac n E > 0; \ x^{(2)} =   \frac n E  + \frac V E \sqrt {\frac n E }  > x^{(1)}; \\
x^{(3)} &=&  \frac n E + \frac V E \sqrt {\frac n E + \frac V E \sqrt {\frac n E } }  > x^{(2)}, ... \nn
\eneq
and so on. Therefore,
we can see that $x^{(k)}$ is a monotonically increasing
sequence bounded by $U_1$, thus the algorithm is convergent.

In the multidimensional case we were not able to prove the convergence rigorously.
However, numerical calculations for many scenarios show that if we set
some stopping rule, e.g. stop the sequential
algorithm when the difference in global costs is less than $0.5$,
then the number of iterations
does not exceed 10 - 15 steps.

As a result, for any feasible probability $P$ this step-wise optimization algorithm
calculates
the optimal site's allocation satisfying conditions of optimization problem
with the optimal cost.

\subsubsection{Numerical example}\label{sec513}

Consider an artificial case study which by the design is very similar to real studies.

In this study it is planned to recruit 250 patients during 2 years.
There are 16 countries where all sites in each country are planned to be activated
in the interval between 30 and 210 days.

The first four columns in the Table 1 describe the enrollment design for this study.
The columns "Low" and "Upp"
reflect the vectors $\bar H$ and $\bar U$ of the lower and upper bounds
for the number of sites
in condition (\ref{sites-bounds}). The column "Rate" shows the mean monthly
enrollment rate for each site
in a corresponding country. The column "Cost" shows the cost in USD for
one patient enrolled in each country.
It is assumed that the coefficient of variation of the enrollment rates is
the same and equal to 1.2, which corresponds
to the medium variation, and assumed that the costs per including one site
are the same in all countries
and equal to \$5000.

%\small{
{\small
\begin{center}
\begin{tabular}{|l|c|c|c|c|c|c|c|c|c|}
\hline
 & Low & Upp & Rate & Cost  & Opt.alloc.  & Opt.alloc. & Opt.alloc. & Opt.alloc. & Opt.alloc.  \\
\hline
Country $ \backslash$ PoS &  &  &  &  & 0.5 & 0.6 & 0.7 & 0.8 &  0.9  \\
\hline
Country1 & 0 & 7 & 0.42  & 15600 & 0 & 0 & 0 & 0 & 0  \\
\hline
Country2 & 0 & 4 & 0.43 & 14250 & 0 & 0 & 0 & 0 &  1  \\
\hline
Country3 & 2 & 5 & 0.22 & 13550 & 2 & 4 & 5 & 5 &  5  \\
\hline
Country4 & 0 & 4 & 0.55 & 14200 & 3 & 4 & 4 & 4 &  4  \\
\hline
Country5 & 0 & 6 & 0.3 & 13800 & 6 & 6 & 6 & 6 &  6  \\
\hline
Country6 & 1 & 7 & 0.57 & 14300 & 1 & 1 & 2 & 4 &  6  \\
\hline
Country7 & 1 & 5 & 0.21 & 13400 & 5 & 5 & 5 & 5 &  5  \\
\hline
Country8 & 1 & 7 & 0.25 & 14250 & 1 & 1 & 1 & 1 &  1  \\
\hline
Country9 & 2 & 5 & 0.16 & 12300 & 5 & 5 & 5 & 5 &  5  \\
\hline
Country10 & 0 & 7 & 0.19  & 13800 & 1 & 1 & 0 & 0 &  0  \\
\hline
Country11 & 2 & 7 & 0.18 & 14600 & 2 & 2 & 2 & 2 &  2  \\
\hline
Country12 & 2 & 7 & 0.62 & 16380 & 2 & 2 & 2 & 2 &  2  \\
\hline
Country13 & 0 & 4 & 0.45 & 13400 & 4 & 4 & 4 & 4 &  4  \\
\hline
Country14 & 0 & 5 & 0.23 & 11200 & 5 & 5 & 5 & 5 &  5  \\
\hline
Country15 & 0 & 5 & 0.3 & 14000 & 1 & 1 & 1 & 1 &  1  \\
\hline
Country16 & 2 & 7 & 0.39 & 14100 & 2 & 2 & 2 & 2 &  2  \\
\hline
Total & 14 & 92 & - & - & 40 & 43 & 45 & 46 &  49  \\
\hline
Opt Cost & - & - & - & - & 3,643,470 & 3,902,851 & 4,135,948 & 4,415,110 &  4,879,621  \\
\hline
\end{tabular}
%{\bf Table 1.} Optimal sites' allocation.
\end{center}
}

\begin{center}
{\bf Table 1.} Optimal sites' allocation.
\end{center}

Using the approach proposed in Section \ref{linear}, it is possible to solve
"Optimization problem 2" and for
a given range of PoS calculate the optimal allocations of sites in these countries.

The columns named "Opt.alloc" in the Table 1
show for each target PoS in the range 0.5, 0.6,..,0.9, the optimal allocation of
sites in these countries such that the corresponding PoS will be reached
with minimal total cost.

The last row "Opt Cost" shows the total cost of study design including patients
and sites costs for each
optimal allocation.

For example, in "Country1" there is rather high cost for patients, so
it's not efficient to include sites
from this country.
On contrary, in "Country5" the cost is not that high and there is a medium mean rate.
Thus, the optimization shows that
this country is more preferable and it is cost-efficient to include
all 6 sites (out of max 6)
in the study design.

The dimension of this problem is $7.11 \times 10^{12}$, so
this problem cannot be solved by using a method of
direct search which is proposed in the next Section \ref{sec514}
for studies with not so large
number of countries.

\subsubsection{The number of countries is not so large}\label{sec514}

If the number of countries is not so large, we can use the direct search.
Consider a general setting in "Optimization problem 1".

If there are no restrictions on the enrollment, the global cost
$C(T, \bar N, \bar C, \bar c, \bar Q) $
is the sum of costs defined by relations (\ref{cost1})-(\ref{cost3}),
where for accelerating
computations we represent the cost for patients in the linear form (\ref{cost2-2}).

Using Lemma \ref{Lem1}, for any given allocation of sites $\bar N$,
we can approximate the distribution of the global enrollment
process at time $T$, $n(T)$,
by the distribution of a PG random variable
$PG(A(T,\bar N),B(T,\bar N))$, where
according to (\ref{A-B-Glob}),
\bee{A-B-Glob2}
A(T,\bar N) = E^{2}(T,\bar N) / S^{2}(T,\bar N), \
B(T,\bar N) = E(T,\bar N) / S^{2}(T,\bar N)
\ene
and the functions $E(T,\bar N)$ and $ S^{2}(T,\bar N)$,
using relations (\ref{mean-time-site}), (\ref{Vs}),
are calculated according to (\ref{MeanGlob}) as
\bee{ETN-2}
E(T,\bar N) = \sum_{s=1}^S N_s m(s) R(s); \
S^{2}(T,\bar N) = \sum_{s=1}^S N_s \si^2(s) V(s)
\ene

Therefore, a function $P(n,T,\bar N)$ in (\ref{PoS-0})
based on the results of Section \ref{sec2-3} is
\bee{Feasible-2}
P(n,T,\bar N) = 1 - \Pr( PG(A(T,\bar N),B(T, \bar N)) \le n-1)
\ene

Correspondingly, the probability $P$ is feasible (can be reached for some allocation)
if
\bee{PoS-0-2}
1 - \Pr( PG(A(T,\bar U),B(T, \bar U)) \le n-1) \ge P
\ene
where $\bar U$
is the vector of upper bounds
for the number of sites in countries.

Note that the representation (\ref{ETN-2}) in the form of linear dependence
on the vector $\bar N$
substantially accelerates computations, the values $R(s)$ and $V(s)$
can be calculated in advance and
then on each step we use only dependence on $\bar N$.

The recurrent step-by-step algorithm (complete search)
is designed as follows.

Denote by $W_2$ a set of all possible allocations of vector $\bar N$
given restrictions (\ref{sites-bounds}). The dimension of this set is
\bee{Dimen}
Dim = \prod_{s=1}^S (U_s - H_s +1)
\ene

Let us consider any recurrent algorithm that can choose on step $k$
some allocation $\bar N_k$ without repetition in such a way that the set
$\{ \bar N_k, k = 1,..,Dim \}$
coincides with the set $W_2$ and $\bar N_1 = \bar H$, $\bar N_{Dim} = \bar U$.

For ease of notation denote $P(\bar N)= P(n,T,\bar N)$;
$C(\bar N) = C(T, \bar N, \bar C, \bar c, \bar Q)$.
Let us take a desirable feasible probability $P$ to complete enrollment in time.
Consider the following recurrent procedure.

Introduce the target vector $\bar Z$ and set the initial value
$\bar Z = (C(\bar U),\bar U)$.
Denote the first component of $\bar Z$ as $\bar Z[1]$.

Then on any step $k$,

if $P(\bar N_k) < P$, go to step $k+1$;

if $P(\bar N_k) \ge P$, then check:

if $C(\bar N_k) \ge \bar Z[1]$, go to step $k+1$;

if $C(\bar N_k) < \bar Z[1]$, then set the new value for target vector $\bar Z$:
$\bar Z = (C(\bar N_k),\bar N_k)$, and go to step $k+1$.

Finally,
this algorithm will come to the optimal target vector $\bar Z_{opt}$
where the components $(2,..,S+1)$ define a feasible allocation $\bar N_{opt}$
that satisfies condition $P(\bar N_{opt}) \ge P$
with minimal cost $C(\bar N_{opt})$.

Computations for different scenarios show that using R, for $Dim = 10^9$
the time of calculation is about 60 min.

For example, for a study with 12 countries and variation in every country about 5 sites,
the time of calculation is about 15 min, which suits practical purposes.

Therefore, the problem to find an optimal enrollment design for unrestricted enrollment
can be efficiently solved, as for studies with not so many countries (up to 12) we can use
the exact algorithm based on the direct search, and for larger studies we can use the approach
based on the normal approximation of the global enrollment process and step-wise
linearisation recurrent algorithm
using simplex method.

\subsection{Restricted enrollment}\label{Opt-Restr}

In this case we also consider two cases: not so large number of countries and vice versa.

For the case of not so large number of countries we can design an optimization algorithm
based on the direct search using  similar steps as described in Section \ref{sec514}.
However, for restricted enrollment the calculations of PoS are based on rather complicated formulae
using a "distributional approach" which is based on using
the convolution of country restricted processes as described in Section \ref{GlobRest},
and that should be repeated on each step of the recurrent algorithm, which takes longer time.
Therefore, the direct search will work longer and can be realistically applied
to studies with up to 8-10 countries with the range up to 5-8 sites in each country.

When the number of countries is larger,  we can use a normal approximation.

As relation (\ref{time-global})
is also valid for the restricted process,
the same condition (\ref{Crit1}) based on the normal approximation should be satisfied
for the restricted process with country caps in order to
complete enrollment up to time $T$ with probability $P$,
where instead of the global mean $E(T,\bar N)$ and the variance $G^2(T,\bar N)$ we should use the global
values $E(T,\bar N, \bar L)$ and $G^2(T,\bar N, \bar L)$
which are the sums of the means and the variances of restricted processes in countries defined
in (\ref{Ecap2-2}) -  (\ref{VarCap}) and calculated as follows:
$$
E(T,\bar N, \bar L) = \sum_{s=1}^S \E[n^{L(s)}(I_s,t)], \
G^2(T,\bar N, \bar L) = \sum_{s=1}^S \Va[n^{L(s)}(I_s,t)]
$$
Then the distribution of the global restricted process $n(t)$ can be calculated using a normal approximation
with mean $E(T,\bar N, \bar L)$ and variance $G^2(T,\bar N, \bar L)$.

Correspondingly,
PoS can be calculated using (\ref{PoS}) and a normal approximation for $n(T_{plan})$:
\bee{PoS-2}
PoS \approx \Phi \Big(  \frac {E(T_{plan},\bar N, \bar L) - n}
{\sqrt{G^2(T_{plan},\bar N, \bar L)}} \Big)
\ene
where $\Phi(x)$ is a CDF of a standard $\cN(0,1)$ normal distribution.

This approach is computationally simpler and takes much less time for calculations.
For example, for a particular scenario of the study with 40 countries and some allocation of caps,
the time to compute probability of success using relation (\ref{PoS-2}) is 190 times less compared
to using a "distributional approach" discussed in Section \ref{GlobRest}.
This essentially helps to analyze various scenarios in real time.
However, as the normal approximation here is used on a global level,
it is recommended to apply this approximation to trials with at least $10 \div 15$ countries.

To solve the optimization problem for a large number of countries
we can use so called evolution or genetic algorithms, where on each step
the appropriate characteristics of the global process and PoS are calculated
using the normal approximation as described above.

\subsubsection{Evolution algorithms}

Evolution algorithms (Differential Evolution - DE)
were designed as some type of random search algorithms using similarity
with genetic mutations \cite{Du16}.
They belong to the class of genetic algorithms which use biology-inspired operations
of crossover, mutation,
and selection on a population in order to minimise an objective function
over the course of successive generations.

As other evolutionary algorithms, DE solve optimization problems
by evolving a population of candidate solutions using alteration and selection operators.
 DE use
floating-point instead of bit-string encoding of population members,
and arithmetic operations instead
of logical operations in mutation. DE are particularly well-suited
to find the global optimum
of a real-valued function of real-valued parameters, and do not require
that the function be either
continuous or differentiable.

The advantage of these algorithms is that they are suitable for solving large dimensional problems
and can be applied to a general
setting in Optimization
problem 1 (\ref{PoS-0}) where PoS is calculated using the algorithms described in
Section \ref{Opt-Restr} for restricted enrollment.

Note that by nature this is some special form
of random search, thus, the outputs can be different for different runs.
It  may also take a substantial time to calculate the optimal point, and
there is no guarantee
that the output will provide a global optimum.
However, a comparison with the results obtained by using direct search shows
that in all
considered examples the evolution algorithms lead to the same results as the exact
algorithm using direct search.

\section{Conclusions}\label{concl}

A new analytic technique for modeling and predicting
patient enrollment on country level
using the approximation of the enrollment process in a country by a
Poisson-gamma process with aggregated parameters is developed.

A novel analytic technique for modeling the enrollment under some restrictions
(enrollment caps in countries) is also developed.

These techniques form the basis for solving the problem of optimal
trial enrollment design:
find an optimal allocation of sites/countries that minimizes
the total trial cost
under the condition that the probability to reach a planned number
of patients in time
is no less than a given probability.

Different techniques to find an optimal solution for low and high dimensional
optimization problems are proposed.

The developed techniques supported by R-software
have a huge potential for improving the efficiency and quality  of clinical trial operation,
and for cost savings.

\section*{Acknowledgement}
%\begin{acknowledgement}
The authors are thankful to Data Science team at the Center for Design \& Analysis,
Amgen Inc. for
useful discussions and providing data from real clinical studies.
%\end{acknowledgement}

\appendix

\section{Appendix}\label{Append}

\addcontentsline{toc}{section}{Appendix}

\subsection{Approximation of the convolution of PG variables}\label{app1}

Let us provide some numerical calculations
to support the results of Lemma \ref{Lem1}.

Consider a country $I$ with $K$ sites. Assume that the enrollment
rates $\la_i$ in all sites are gamma distributed with the same
parameters $(\a,\be)$. Consider some interim time $t$ and denote by $v_i$
the duration of enrollment in site $i$ up to time $t$.
Then, according to (\ref{CumRate}), the global cumulative enrollment
rate in country $I$ is
$$
\Lambda(I,t) = \sum_{i=1}^K \lambda_i v_i
$$
As noted in Section \ref{sec2-2}, if for all $i \in I$,  $v_i \equiv v$,
then $\Lambda(I,t)$ has the same distribution as $\Gam(\alpha K,\beta/v)$,
and the enrollment process $n(I,t)$ in country $I$
is a PG process which has the same distribution as $PG(v, \alpha K,\beta)$ variable.

However, in realistic cases the enrollment durations $v_i$
are different, $\La(I,t)$ does not have a gamma distribution, so,
 $n(I,t)$ does not have a PG distribution.
Nevertheless, using Lemma \ref{Lem1}, a distribution of $n(I,t)$
can be approximated by a PG-process with some
 parameters.

The accuracy of this approximation was evaluated using numerical calculations
for many different scenarios and the results led to the same conclusions.
Here for the illustration
we provide the analysis using only one example (see \cite{an-aus20}).

Put $\a=1.5, \be=150$.
Assume that $v_i, i=1,..,K$, are taken using a uniform grid in interval $[1,300]$ as:
$round((1:K)*300/K)$. This reflects
a reasonably large variation in $v_i$.
In site  $i$  the distribution of the number of enrolled patients
is calculated
as a vector $pp_i$ of the length $L+1$
using a PG distribution with parameters $(v_i,\a,\be)$
and formula in R:
\begin{verbatim}
dnbinom(0:L,size=alf,prob=be/(be+v[i]))
\end{verbatim}

Consider the length $L=50$ as for $k > 50$
the probabilities to enrol $k$ patients are zeros up to 4 digits.

The probability distribution of the enrollment process in the country
is a convolution of probability distributions $pp_i$ and can be
calculated numerically using very fast procedure in R based on
function $convolve()$.
Denote the resulting distribution
by $ppK$. This is the exact distribution up
to the accuracy of computations.

We can also approximate the probability distribution of
the enrollment process in the country
by a PG distribution using relations (\ref{e20}),(\ref{e21}) and Lemma \ref{Lem1}.
Denote the approximative distribution by $ppKPG$.

\begin{figure}\label{Fig3}
\centering
\includegraphics[width=14.0cm,height=7cm]{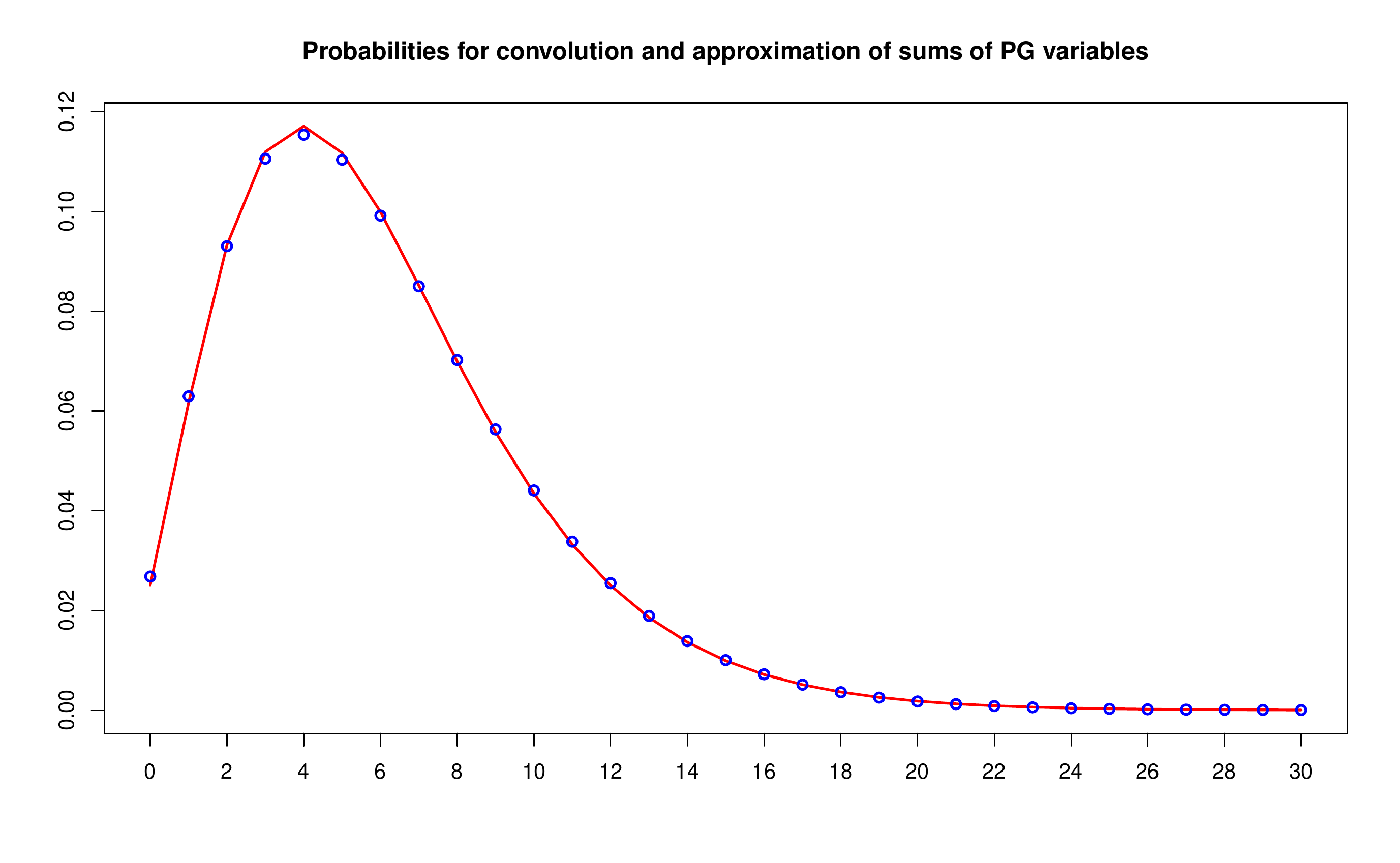}
\caption{
Approximation of the distribution of the enrollment process in the country
with three sites by the distribution of a PG variable with aggregated parameters.
A continuous line shows the exact values of the distribution of the enrollment process
calculated using a convolution of PG processes in sites,
dotted line shows its approximation by a PG distribution.
}
\end{figure}

The computations show
that probability distributions are very close even
for small $K=2,3$, where the absolute difference
$$
Dif(K) = \max( \Big | ppK[i]-ppKPG[i] \Big |, i=0,..,L)
$$
is decreasing when $K$ is increasing.

The table below describes the values of $Dif(K)$ for different $K$.

{\small
%{\footnotesize
\begin{center}
\begin{tabular}{|l|c|c|c|c|c|c|c|}
\hline
K  & 2 & 3  & 5 & 8 & 10 & 15 & 20  \\
\hline
Dif(K) & 0.0019 & 0.0017 & 0.0011 & 0.00075 & 0.00059 & 0.00039 &  0.00029  \\ \hline
\end{tabular}
\end{center}
}

The plot in Figure 1
illustrates the case $K=3$.
%where a continuous line shows the exact values of the distribution
%of the enrollment process in the country and the dotted line shows
%the approximative PG distribution.
Here the difference between the exact and approximative probability distributions is negligible.
Thus, the result of Lemma \ref{Lem1}
can be used very efficiently in practice.

\subsection{Calculation of the mean of the restricted process in one site}\label{App-mean}

Consider a site $i$ where the enrollment process $n_i(t)$ is restricted by cap $L$.
Relation (\ref{e2}) implies that
\beeq{mean}
\E[n_i^{L}(t)] = \sum_{k=0}^{L-1} k P(k,t,u) + L (1- \Pr(PG(\a,\be,x(t,u)) \le L-1))
\eneq
Denote the first sum in the right-hand side as $S_1$.
To simplify calculations, we use $t$ instead of $x(t,u)$.
Using the change of variable ($k-1 \to k$) and relation
$\Ga(\a+1)=\a\Ga(\a)$, we get
\beeq{sumS1}
S_1 &=& \sum_{k=0}^{L-1} k\frac{\Ga(k+\a)} {k! \, \Ga(\a)}
\frac {\be^\a t^k}{(\be+ t)^{\a+k}} \nn \\
&=& \frac {\a t}{\be} \sum_{k=1}^{L-1} \frac{\Ga(k-1+\a+1)} {(k-1)! \, \Ga(\a+1)}
\frac {\be^{\a+1} t^{k-1}}{(\be+ t)^{\a+1+k-1}}
 \\
&=& \frac {\a t}{\be} \sum_{k=0}^{L-2} \frac{\Ga(k+\a+1)} {k! \, \Ga(\a+1)}
\frac {\be^{\a+1} t^{k}}{(\be+ t)^{\a+1+k}}  \nn
\\
&=& \frac {\a t}{\be} \Pr(PG(\a+1,\be,t)\le L-2) \nn
\eneq

Finally,
putting back $t = x(t,u)$, we get relation (\ref{Ecap2-1}).

\subsection{Calculation of the 2nd moment of the restricted process in one site}\label{App-2nd}

Relation (\ref{e2}) implies that
\beeq{2ndmoment}
\E[(n_i^{L}(t))^2] = \sum_{k=0}^{L-1} k^2 P(k,t,u) + L^2 (1- \Pr(PG(\a,\be,x(t,u)) \le L-1))
\eneq

To simplify calculations, we use again $t$ instead of $x(t,u)$.
Denote the first sum in the right-hand side as $M_2$.
Consider first an auxiliary sum
$$
A = \sum_{k=0}^{L-1} k(k-1) \frac{\Ga(k+\a)} {k! \, \Ga(\a)}
\frac {\be^\a t^k}{(\be+ t)^{\a+k}}
$$
By definition of $S_1$ in (\ref{sumS1}) we see that \
$
M_2=A+S_1.
$
Then, using the change of variable ($k-2 \to k$) and relation
$\Ga(\a+2)=  \a(\a+1)\Ga(\a)$, we get
\beeqn
A &=& \sum_{k=0}^{L-1} k(k-1) \frac{\Ga(k+\a)} {k! \, \Ga(\a)}
\frac {\be^\a t^k}{(\be+ t)^{\a+k}} \\
&=& \frac {\a(\a+1) t^2}{\be^2} \sum_{k=2}^{L-1} \frac{\Ga(k-2+\a+2)} {(k-2)! \, \Ga(\a+2)}
\frac {\be^{\a+2} t^{k-2}}{(\be+ t)^{\a+2+k-2}}
\\
&=& \frac {\a(\a+1) t^2}{\be^2} \sum_{k=0}^{L-3} \frac{\Ga(k+\a+2)} {k! \, \Ga(\a+2)}
\frac {\be^{\a+2} t^{k}}{(\be+ t)^{\a+2+k}} \\
&=&  \frac {\a(\a+1) t^2}{\be^2} \Pr(PG(\a+2,\be,t)\le L-3)
\eneqn

Thus,
\beeqn
M_2 &=&  \frac {\a(\a+1) t^2}{\be^2} \Pr(PG(\a+2,\be,t)\le L-3) \\
&+& \frac {\a t}{\be} \Pr(PG(\a+1,\be,t)\le L-2)
\eneqn

Finally, putting back $t = x(t,u)$, we get relation (\ref{Cap2-1}).


\begin{thebibliography}{99.}%
% Use the following syntax and markup for your references if
% the subject of your book is from the field
% "Mathematics, Physics, Statistics, Computer Science"

\bibitem{anfed05}
V. Anisimov and V. Fedorov.
Modeling of enrolment and estimation of parameters in multicentre  trials,
{\em GSK BDS Technical Report 2005-01}, 33p, 2005.

\bibitem{anfed07a}
V. Anisimov and V. Fedorov. Design of multicentre clinical
trials with random enrolment. In {\em "Advances in Statistical
Methods for the Health Sciences"}.
Applications to Cancer and AIDS Studies,
Genome Sequence Analysis, and Survival Analysis",
Series: Statistics for Industry and Technology, Balakrishnan N.;
Auget J.-L.; Mesbah M.; Molenberghs G. (Eds.) Birkhauser.
Ch.25:387--400, 2007.

\bibitem{anfed07}
V. Anisimov  and V. Fedorov.  Modeling, prediction and adaptive
adjustment of recruitment in multicentre trials, {\em Statistics in Medicine},
26, 27, 4958--4975, 2007.

\bibitem{an-dow-fed07}
V. Anisimov, D. Downing  and  V. Fedorov. Recruitment in multicentre trials:
prediction and adjustment,
{\em mODa 8 - Advances in Model-Oriented Design and Analysis}, 1--8, 2007.
%Lopez-Fidalgo J,
%Rodriguez-Diaz JM, Torsney B (Eds), Physica-Verlag 1--8.

\bibitem{an08}
V. Anisimov. Using mixed Poisson models in patient recruitment in multicentre clinical trials,
{\em Proc. of the World Congress on Engineering}, II, 1046--1049, 2008.


\bibitem{an09a}
V. Anisimov. Predictive modelling of recruitment and drug
supply in multicenter clinical trials. In: {\em Proc. of the Joint
Statistical Meeting}, Biopharmaceutical Section, Washington, DC,
American Statistical Association. 1248--1259, 2009.


\bibitem{an11a}
V. Anisimov. Statistical modeling of clinical trials (recruitment and randomization),
{\em Communications in Statistics - Theory and Methods}, 40, 19-20, 3684--3699, 2011.


\bibitem{an11b}
V. Anisimov. 	Predictive event modelling in multicentre clinical trials with waiting
time to response, {\em Pharmaceutical Statistics}, 10, 6, 517--522, 2011.

\bibitem{an12}
V. Anisimov.
Discussion on the paper 'Prediction of accrual closure date in multi-center
clinical trials with discrete-time Poisson process models'
by G. Tang, Y. Kong, C. Chang, L. Kong, and J. Costantino,
{\em Pharmaceutical Statistics}; 11, iss. 5:357--358, 2012.


\bibitem{an16a}
V. Anisimov. 	 Predictive hierarchic modelling of operational characteristics
in clinical trials, {\em Communications in Statistics - Simulation and Computation},
45, 5, 1477--1488, 2016.

\bibitem{an16b}
V. Anisimov.  Discussion on the paper "Real-time prediction of clinical trial enrollment
and event counts: a review" by D.F. Heitjan et al. {\em Contemporary Clinical Trials},
40, 7--10, 2016.


\bibitem{an20}
V. Anisimov. Modern analytic techniques for predictive modelling of
clinical trial operations, {\em Quantitative Methods in
Pharmaceutical Research and Development: Concepts and Applications},
Springer International Publ., 361--408, 2020.

\bibitem{an-aus20}
V. Anisimov  and M. Austin.
Centralized statistical monitoring of clinical trial enrollment performance,
{\em Communications in Statistics - Case Studies and Data Analysis}, 6, 4,
392--410, 2020.

\bibitem{an-ArXiv21}
V. Anisimov, S. Gormley, R. Baverstock, and C. Kineza.
Advanced models for predicting event occurrence in event-driven clinical trials
accounting for patient dropout, cure and ongoing recruitment,
{\em arXiv:2108.09196}, 1--17, 20 Aug 2021.

\bibitem{baksenn13}
 A. Bakhshi, S. Senn and A. Phillips. Some issues in predicting patient
  recruitment in multi-centre clinical trials. {\em Statistics in
  Medicine}, 32(30):5458--5468, 2013.

\bibitem{barnard10}
K.D. Barnard, L. Dent  and  A. Cook. A systematic review of models to predict
recruitment to multicentre clinical trials,
{\em BMC Medical Research Methodology}, 10, 63, 2010.

  \bibitem{bates52}
GE. Bates and J. Neyman. Contributions to the theory of
accident proneness, {\em University of California Publications in
Statistics}, 1(9):215--254, 1952.


\bibitem{bernardo04}
J.M. Bernardo  and  A.F.M. Smith. {\em Bayesian Theory}, John Wiley \&  Sons:
Hoboken, NJ, USA, 2004.

  \bibitem{carter05}
R.E. Carter, S.C. Sonne and K.T. Brady.
Practical considerations for  estimating clinical trial accrual periods: Application to a
  multi-center effectiveness study, {\em BMC Medical Research
  Methodology},  5:11--15, 2005.

\bibitem{Du16}
K.-L. Du and  M. N. S. Swamy. {\em Search and Optimization by
Metaheuristics: Techniques and Algorithms Inspired by Nature.}
Birkhauser, Basel, 2016

\bibitem{gajew-sim08}
BJ. Gajewski, SD. Simon and SE. Carlson.
  Predicting accrual in   clinical trials with Bayesian posterior predictive distributions.
  {\em Statistics in Medicine}, %2008;
  27:2328--2340, 2008.

\bibitem{gkioni19}
E. Gkioni, R. Riusd, S. Dodda and C. Gamblea.
A systematic review describes models for recruitment prediction at the
design stage of a clinical trial, {\em Journal of Clinical
Epidemiology}, 115:141--149, 2019.

  \bibitem{heitjan15}
D.F. Heitjan, Z. Ge  and  G.S. Ying.  Real-time prediction of clinical trial
enrollment and event counts: a review, {\em Contemporary Clinical  Trials},
45, part A, 26--33, 2015.

\bibitem{john-kotz93}
NL. Johnson, S. Kotz and AW Kemp. {\em Univariate Discrete
Distributions},
  2nd Ed., John Wiley \& Sons: New York, 1993.

  \bibitem{savy12}
  G. Mijoule, S. Savy and N. Savy. % 2012.
  Models for patients' recruitment in  clinical trials and sensitivity analysis,
  {\em Statistics in Medicine},
  31(16):1655--1674, 2012.


\bibitem{savy17}
MN. Minois, V. Lauwers-Cances, S. Savy, M. Attal, S. Andrieua,
V. Anisimov and  N. Savy. Using Poisson-gamma model to
evaluate the duration of
  recruitment process when historical trials are available.
  {\em Statistics in Medicine}, %2017;
  36(23):3605--3620, 2017.

\bibitem{satt46}
FE. Satterthwaite. An approximate distribution of estimates
of variance components,
{\em Biometrics Bulletin}, %1946;
2, No. 6:110--114, 1946.


\bibitem{senn97}
S. Senn. {\em Statistical Issues in Drug Development}. Wiley:  Chichester, 1997.

\bibitem{senn98}
S. Senn. Some controversies in planning and analysis  multi-center trials.
{\em Statistics in Medicine}, 17, 1753--1756, 1998.

\bibitem{tufts13}
Tufts.  CSDD impact report - 89\% of trials meet enrolment,
but timelines slip, half of sites underenrol, {\em Tufts Center for the
Study of Drug Development, Impact report.} v. 15 (1), 2013.


\bibitem{welch47}
BL. Welch The generalization of Student's problem when
several different population variances are involved.
{\em Biometrika}, 34:28--35, 1947.


\bibitem{williford87}
WO.  Williford, SF. Bingham, DG. Weiss, JF. Collins, KT. Rains and WF. Krol.
The 'constant intake rate' assumption in interim recruitment
  goal methodology for multicenter clinical trials.
  {\em J. Chronic Dis.} %1987:
  40:297--307, 1987.

\end{thebibliography}
\end{document}